\documentclass{raa}
\usepackage{graphicx,times}
\usepackage{subfigure}
\usepackage{natbib}
\usepackage{amssymb,amsmath}
\bibpunct{(}{)}{;}{a}{}{,}

\usepackage{color}

\usepackage[a4paper=true,dvipdfm=true,pagebackref=true]{hyperref}
\hypersetup{pdftitle = Global Mg/Si and Al/Si Distributions on Lunar Surface Derived from Chang'E-2 X-ray Spectrometer, pdfauthor = Dong Wu-Dong, pdfsubject= The subject, pdfkeywords = X-ray fluorescence  Chang'E-2 spacecraft  Moon  Mg/Si and Al/Si composition map}
\hypersetup{colorlinks = true, linkcolor = green, anchorcolor = red, citecolor = blue, filecolor = red, pagecolor = red, urlcolor = red}

\begin{document}

   \title{Global Mg/Si and Al/Si Distributions on Lunar Surface Derived from Chang'E-2 X-ray Spectrometer
$^*$
\footnotetext{\small $*$ Supported by the Science and Technology Development Fund of Macao and the Key Research Program of the Chinese Academy of Sciences.}
}

 \volnopage{ {\bf 2015} Vol.\ {\bf X} No. {\bf XX}, 000--000}
   \setcounter{page}{1}

   \author{Wu-Dong Dong\inst{}, Xiaoping Zhang\inst{}, Meng-Hua Zhu\inst{}, Aoao Xu\inst{},  Zesheng Tang\inst{}}

   \institute{ Lunar and Planetary Science Laboratory, Macau University of Science and Technology, Macao, China; {\it Corresponding authors: xpzhang@must.edu.mo (X. Zhang) and dowudong@gmail.com (W.-D. Dong)}\\
\vs \no
}

\abstract{X-ray fluorescence remote sensing technique plays a significant role in the chemical compositions research of the Moon. Here we describe the data analysis method for China's Chang'E-2 X-ray spectrometer (CE2XRS) in detail and present the preliminary results: the first global Mg/Si and Al/Si maps on the lunar surface.  Our results show that  the distributions of Mg/Si and Al/Si correlate well with the terrains  of the Moon.  The higher Mg/Si ratio corresponding to the mare regions while the lower value corresponding to the highland terrains. The map of Al/Si ratio shows a reverse relationship with the map of Mg/Si ratio.
}

   \authorrunning{W.-D. Dong et al. }            
   \titlerunning{Global Mg/Si and Al/Si Distributions on Lunar Surface Derived from CE2XRS}  
   \maketitle


%
\section{Introduction}           
\label{sect:intro}

Chemical compositions on the lunar surface are  critical in the determination of geochemical nature of lunar terrains and the geologic evolution history  of the Moon. X-ray spectroscopy is considered as one of the most effective  method among all remote sensing techniques to study the major elements abundance on the airless planetary surface, such as the Moon, Mercury, and asteroids (\citealt{Adler+etal+1973}; \citealt{Clark+Trombka+1997}; \citealt{Trombka+etal+2000}; \citealt{Grande+etal+2003}; \citealt{Sun+etal+2008}; \citealt{Nittler+etal+2011}). In this technique, the solar X-rays interact with the  planetary's surface materials producing characteristic X-ray fluorescences at the uppmost  surface layer.  The characteristic fluorescent X-rays of major elements can be detected by  the instrument in orbit , such as K$\alpha$ lines of Mg (1.254 keV), Al (1.487 keV), Si (1.740 keV), Ca (3.692 keV), Ti (4.511 keV), and Fe (6.404 keV).

X-ray spectrometer was carried on many  spacecrafts in the past. The X-ray spectrometer onboard the Russian Luna 12 orbiter sent in 1966 first successfully observed fluorescent X-rays from the Moon (\citealt{Mandel'Shtam+etal+1968}). Later in 1971 - 1972, X-ray fluorescences were measured  by proportional counters in lunar equatorial region of the nearside in Apollo 15 and 16 missions,  covering $\sim$ 10 $\%$  of  the global lunar surface.  This observation was considered as the only observation of the large-scale chemical compositions of lunar surface from the X-ray spectrometer. The Al/Si and Mg/Si ratios were derived from Apollo X-ray Spectrometer data,  providing a preliminary understanding of the local distribution feature of the lunar surface chemical composition (\citealt{Adler+etal+1972a, Adler+etal+1972b}; \citealt{Clark+1979}). The Demonstration of a Compact Imaging X-ray Spectrometer  (D-CIXS) payload onboard SMART-1 launched by ESA in 2003 detected the characteristic X-ray lines from rock-forming elements during several solar flare events (\citealt{Grande+etal+2003}; \citealt{Grande+etal+2007}; \citealt{Swinyard+etal+2009}).  Unfortunately, the detector  suffered severe radiation in  orbit during  its observation,  from which no accurate quantitative analysis was presented. The X-ray spectrometer onboard  Kaguya  spacecraft that was launched in 2007  also suffered  the same problem  (\citealt{Yamamoto+etal+2008}; \citealt{Okada+etal+2009}). The X-ray spectrometers onboard the first Chinese lunar spacecraft Chang'E-1 launched in 2007 (\citealt{Sun+etal+2008}; \citealt{Ouyang+etal+2008}; \citealt{Ouyang+etal+2010a, Ouyang+etal+2010b}) and Indian Chandrayaan-1 launched in 2008 (\citealt{Grande+etal+2009}) successfully detected the fluorescent X-rays from rock-forming elements (\citealt{Ouyang+etal+2008}; \citealt{Peng+2009}; \citealt{Narendranath+etal+2011}; \citealt{Weider+etal+2012, Weider+etal+2014}). Since the Sun was in quiescent period at that time, the solar flares were few and the incident X-ray intensity was not powerful enough. As a result, the  data  is not sufficient for producing a global elemental distribution map.

The Chang'E-2 (CE-2) spacecraft, which was the second unmanned China's lunar probe, was successfully launched on 1 October 2010. An innovative X-ray spectrometer (CE2XRS) was onboard  CE-2.  \cite{Ban+etal+2014} derived the elemental abundances of Mg, Al, Si, Ca and Fe in the lunar Oceanus Procellarum region using CE2XRS data acquired during an M-class solar flare event. However, there is no global elements abundance map of the Moon from X-ray spectrometer until now. Fortunately, the Sun was in  its active period  in which a number of solar flare events  were observed during the CE-2 mission . After  more than half year's observation, large  amounts of scientific data  were obtained by CE2XRS. It is, therefore, the best opportunity to derive the first global chemical composition maps of major elements with CE2XRS.

In this paper, we describe the data analysis procedure and present the global Mg/Si and Al/Si distributions on lunar surface derived from CE2XRS data. This is the first global map  derived by the remote X-ray fluorescence spectroscopy on the Moon. Lunar global geochemical features are discussed based on the element distribution maps. Three major geological units on the Moon (lunar mare, highland and South Pole-Aitken basin) are identified.

\section{Chang'E-2 X-ray Spectrometer}
\label{sect:Obs}
 The X-ray spectrometer, one of the scientific payloads onboard CE-2 spacecraft,  was aimed to obtain major rock-forming elements  distributions (i.e., Mg, Al, Si, Ca, Ti, and Fe) on the Moon. Since the  detailed structure and specification of CE2XRS has been  presented by \cite{Peng+etal+2009a} and \cite{Ban+etal+2014},  we here only give a brief description.

\begin{figure}
   \centering
   \includegraphics[width=10.0cm, angle=0]{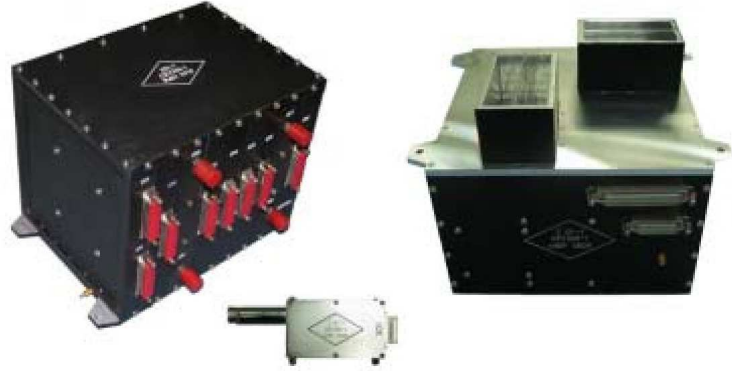}
   \caption{The CE2XRS instruments, from left to right listed are electronic box, solar X-ray monitor, lunar X-ray detector. The photograph is from \cite{Peng+etal+2009a}.}
   \label{Fig_inst}
   \end{figure}

The CE2XRS was designed as a compact detector. It consisted of lunar X-ray detector, solar X-ray monitor and electric box (see Figure~\ref{Fig_inst}), which was an improved one compared to the X-ray spectrometer on Chang'E-1 spacecraft. A pair of orthogonal collimators made of permanent magnet limits the field of view of the detectors to make the best spatial resolution of $\sim$ 200 km $\times$ 70 km at a distance of 100 km from lunar surface. The layout of the detectors and collimators were described by \cite{Peng+etal+2009b}. The lunar X-ray detector was designed to measure the lunar fluorescent X-rays. It  was composed of a soft X-ray detector array  SXDA and a hard X-ray detector array  (HXDA). The SXDA had 4 Si-PIN detectors named 1111-9, 1111-10, 2222-9, 2222-10 . The detective energy range of SXDA was 0.5-10 keV and the energy resolution was 300 eV@5.95 keV.  As for HXDA, the detective energy range was 25-60 keV and the energy resolution was 6.0 keV@59.5 keV. The solar X-ray monitor (SXM) had a small Si-PIN detector . It was the same as  the detector in SXDA except the  effective detector area.  The SXM was designed to measure the solar X-ray spectrum simultaneously. A $^{55}$Fe radioactive source  emplaced on the  1111-9 detector in SXDA was used to calibrate the instrument in  orbit. X-rays emitted from the $^{55}$Fe source are the Mn K$\alpha$ and K$\beta$ lines at 5.898 keV and 6.490 keV, respectively. The instrument specifications of the X-ray spectrometer are listed in Table~\ref{tab1}.

\begin{table}[h]
\bc
\begin{minipage}[]{100mm}
\caption[]{Technical specification of CE2XRS from \cite{Peng+etal+2009a}.\label{tab1}}
\end{minipage}
\setlength{\tabcolsep}{16pt}
\small
 \begin{tabular}{cccc}
  \hline\noalign{\smallskip}
Component & SXD & HXD & SXM\\
  \hline\noalign{\smallskip}
Detector & Si-PIN (4 chips) & Si-PIN (16 chips) & Si-PIN (1 chip)\\
Filter & 12.5 $\mu$m Be & $1$ $\mu$m Al & 12.5 $\mu$m Be\\
Effective area & 1 cm$^2$ & 16 cm$^2$ & 0.2 mm$^2$\\
Detective range & 0.5-10 keV & 25-60 keV & 0.5-10 keV\\
Energy resolution & 300 eV@5.9 keV & 6 keV@59.5 keV & 300 eV@5.9 keV\\
  \noalign{\smallskip}\hline
\end{tabular}
\ec
\end{table}

\section{CE2XRS Data Analysis}
The CE-2 spacecraft kept a polar orbit with a period of $\sim$ 118 minutes at an average altitude of 100 km. There were 2739 orbits data collected from October 15, 2010 to May 20, 2011. In our analysis, we use the CE2XRS level 2C dataset and only analyze the data from the detector 1111-9 in SXDA. The data format is shown in Table~\ref{tab2}. The description of the dataset was presented by \cite{Ban+etal+2014}. A general X-ray spectrometer data analysis method was pubilshed by  \cite{Clark+Trombka+1997} , \cite{Starr+etal+2000} and \cite{Peng+2009}. Here we describe the CE2XRS data analysis method for global elemental distribution map in detail.

\begin{table}[h]
\bc
\begin{minipage}[]{100mm}
\caption[]{The format of CE2XRS level 2C dataset.\label{tab2}}
\end{minipage}
\setlength{\tabcolsep}{16pt}
\small
 \begin{tabular}{cccc}
  \hline\noalign{\smallskip}
Column & Label & Column & Label\\
  \hline\noalign{\smallskip}
1 & time & 7 & instrument\_azimuth\_angle\_2\\
2 & longitude & 8 & solar\_incidence\_angle\\
3 & latitude & 9 & detector\_number\\
4 & distance & 10 & energy\\
5 & instrument\_incidence\_angle & 11 & counts\\
6 & instrument\_azimuth\_angle & 12 & quality\_state\\
  \noalign{\smallskip}\hline
\end{tabular}
\ec
\end{table}

\subsection{Data Check}

\begin{figure}
   \centering
   \includegraphics[width=14.5cm, angle=0]{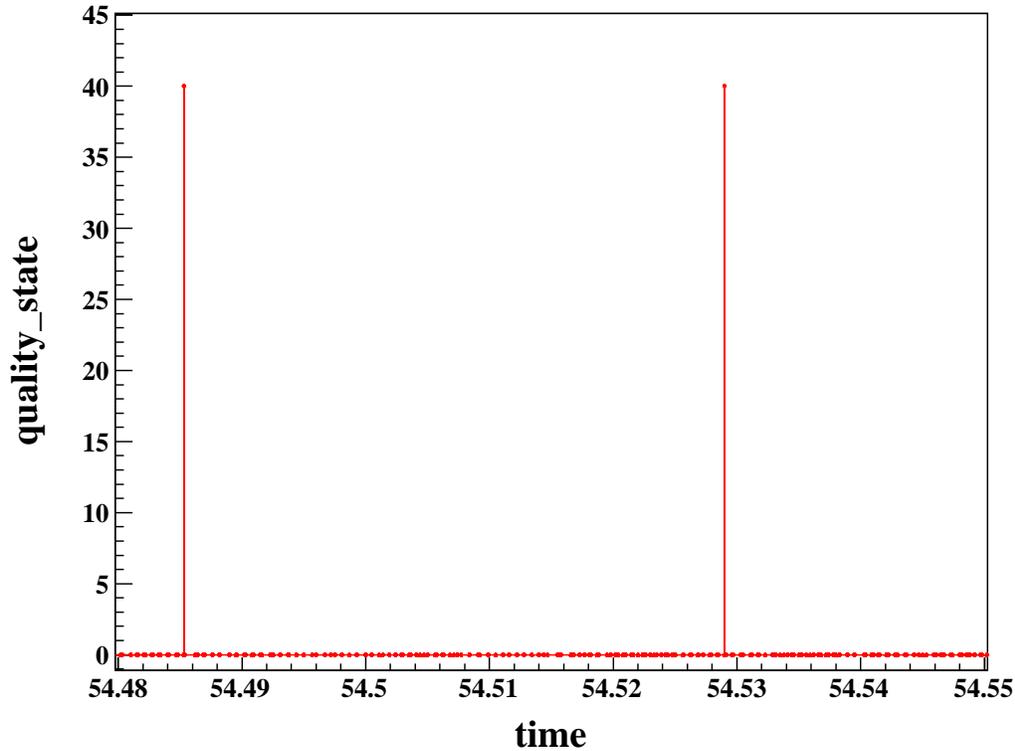}
   \caption{The scatter plot of label ``quality\_state'' (QS) versus time from the data in orbit number 578. The time means the number of days since October 1, 2010. This orbital record started from 11:22 to 13:20 on November 24, 2010 (UTC). The bad record with the ``quality\_state'' ``40'' happened at 11:38 and 12:41. These abnormal records are eliminated in the data check process.}
   \label{Fig_578}
   \end{figure}

\begin{figure}
   \centering
   \includegraphics[width=14.5cm, angle=0]{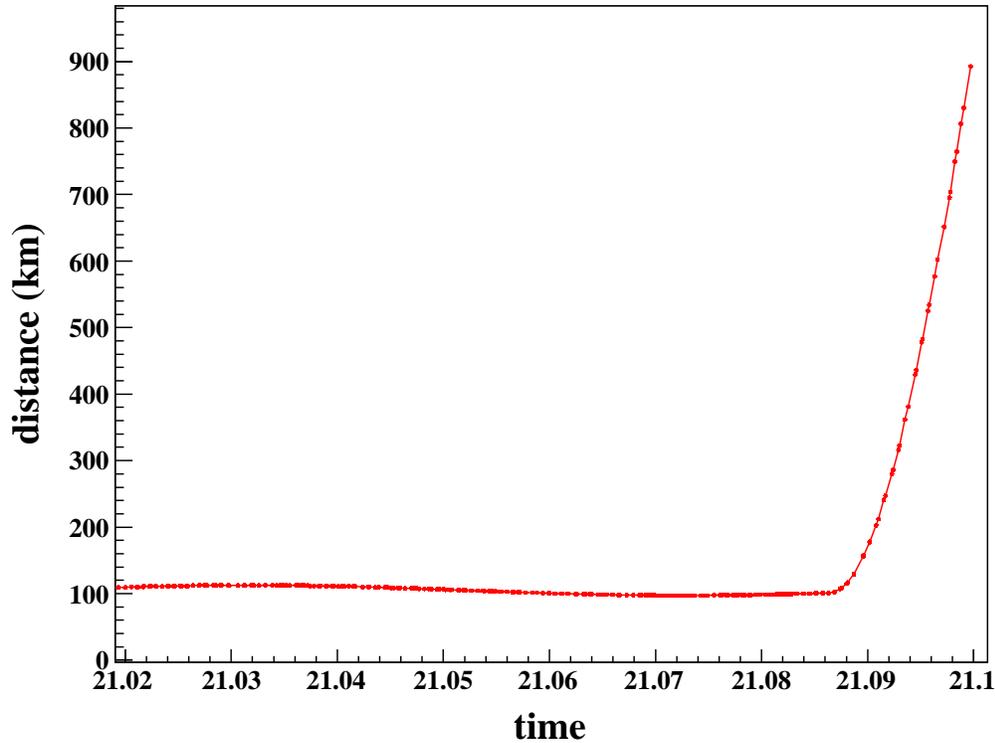}
   \caption{The scatter plot of label ``distance'' versus time from the data in orbit number 169. The time means the number of days since October 1, 2010. This orbital record started from 00:25 to 02:23 on October 22, 2010 (UTC). The ``distance'' (orbit altitude) sharply increased at 02:07. These records are removed while the CE-2 spacecraft might be doing orbital transfer at that time. }
   \label{Fig_169}
   \end{figure}

For the CE2XRS level 2C data, there is a label ``quality\_state'' (see Table~\ref{tab2}) appended to the end of each record. The value of ``quality\_state'' equaled ``00'' (decimal value = 0) indicating that the record is good. If a record label ``quality\_state'' is not equal to ``00'', this record is eliminated (see Figure~\ref{Fig_578}). The label ``distance'' is the altitude between the CE-2 spacecraft and the lunar surface. The normal average work altitude of CE2XRS is $\sim$100 km. The records with the value of label ``distance'' greater than 125 km are deleted (see Figure~\ref{Fig_169}), although the percentage of such measurements is very low.

\begin{figure}
   \centering
   \includegraphics[width=14.5cm, angle=0]{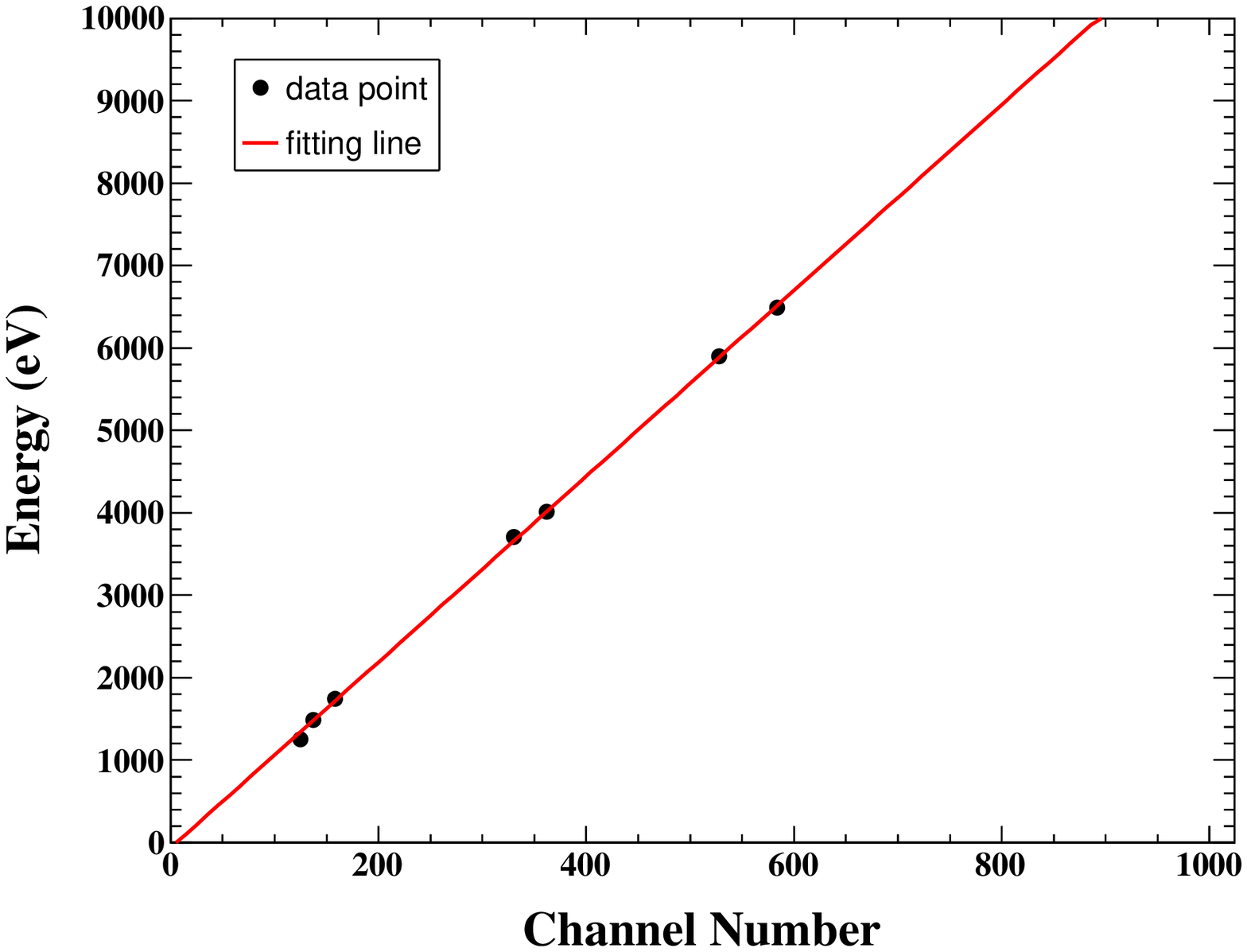}
   \caption{The linear relationship between channel number and energy from detector 1111-9 data. The points from left to right are Mg(K$\alpha$), Al(K$\alpha$), Si(K$\alpha$), Ca(K$\alpha$), Ca(K$\beta$), Mn(K$\alpha$) and Mn (K$\beta$).}
   \label{Fig_ch2eg}
   \end{figure}

Also, we check the relationship between channel number and energy for the detector 1111-9. This detector has 1024 channels. The channel information is stored in the level 2B dataset. The level 2C dataset is generated when the channel-energy calibration was carried out on the level 2B dataset. This preliminary calibration was made by Institute of High Energy Physics and National Astronomical Observatories. We find a good linear relationship between channel number and energy (see Figure~\ref{Fig_ch2eg}) which demonstrates the good quality of the CE2XRS data. After the data check is finished, the final corrected data is ready for spectral analysis.

\subsection{The Solar Activity and Data Selection}
The solar X-ray is the primary excitation source for the fluorescent X-rays generation from lunar surface. When the Sun is in active period, the solar X-ray flux increase significantly. The flux of higher energy X-rays may increase several orders of magnitudes and makes the solar X-ray spectra much harder during solar flare. Hence, fluorescent X-rays from lunar surface increases greatly and the signal-to-noise ratio also goes up in this period. When the Sun is in quiescent period, the fluorescent X-rays is much lower and much more data accumulation time is needed to achieve significant results. The relationship between the variation in solar activity and the intensity of fluorescent X-rays had been derived by \cite{Clark+Trombka+1997}.

In our analysis, the solar plasma temperature ($T_{solar}$) parameter is used to represent the intensity of solar activity. We adopt the solar plasma isothermal model to derive the solar plasma temperature. The solar X-ray flux data are taken from the Geostationary Operational Environmental Satellite (GOES). GOES measures the solar flux in two channels (1-8 \AA{} and 0.5-4 \AA, corresponding energy is 1.55-12.4 keV and 3.1-24.8 keV) simultaneously in every 3s. \cite{Thomas+etal+1985} and \cite{Garcia+1994}  have published methods for calculating solar plasma temperature from GOES data. This model was used by \cite{Trombka+etal+2000} and \cite{Nittler+etal+2001} in the NEAR-Shoemaker XRS papers to interpret the elemental composition of asteroid 433 Eros and also adopted in CE-1 XRS data analysis by \cite{Peng+2009} . The relationship between solar plasma temperature and the solar flux ratio (SFR) was obtained by fitting method  (\citealt{Peng+2009}). The SFR is defined by the ratio of flux in channel 1-8 \AA{} and flux in channel 0.5-4 \AA. When the SFR increases, the solar plasma temperature goes down.

From GOES solar flare event report\footnote{http://www.ngdc.noaa.gov/stp/space-weather/solar-data/solar-features/solar-flares/x-rays/goes/}, we find that there were many solar flare events during the CE2XRS observation period from October 15, 2010 to May 20, 2011. For the solar flare event with level B and above, the footprint of CE2XRS on the Moon covered all of the lunar surface. The solar condition was good for mapping out major elements, such as Mg, Al, Si, on the entire lunar surface. To obtain statistically significant results for globally mapping, the X-ray spectra with the same solar condition and at the same location are accumulated.
To increase the signal-to-noise ratio, only the data acquired in the sunlit side of the Moon with solar incidence angle less than $88^{\circ}$ are accumulated. The final data are categorized by different solar plasma temperatures. We find the amount of data with the solar plasma temperature of 4 MK is sufficient for deriving global distribution map of Mg, Al and Si on the entire lunar surface. Therefore, this solar condition is selected for our analysis. Hereafter, all operations are done on the data at this solar condition.

The lunar surface is partitioned into a series of equal area grids for mapping CE2XRS data. According to the space resolution of CE2XRS and the statistics of the data, the grid size is set to be $20^\circ\times20^\circ$. The partitioned grids start from the lunar equator to the polars with the same latitude intervals but different longitude intervals. The total number of grids is 114. All corrected data in the same grid and in the same solar condition are accumulated to form a spectra for further analysis.

\subsection{Background Determination}
\begin{figure}
   \centering
  \includegraphics[width=14.5cm, angle=0]{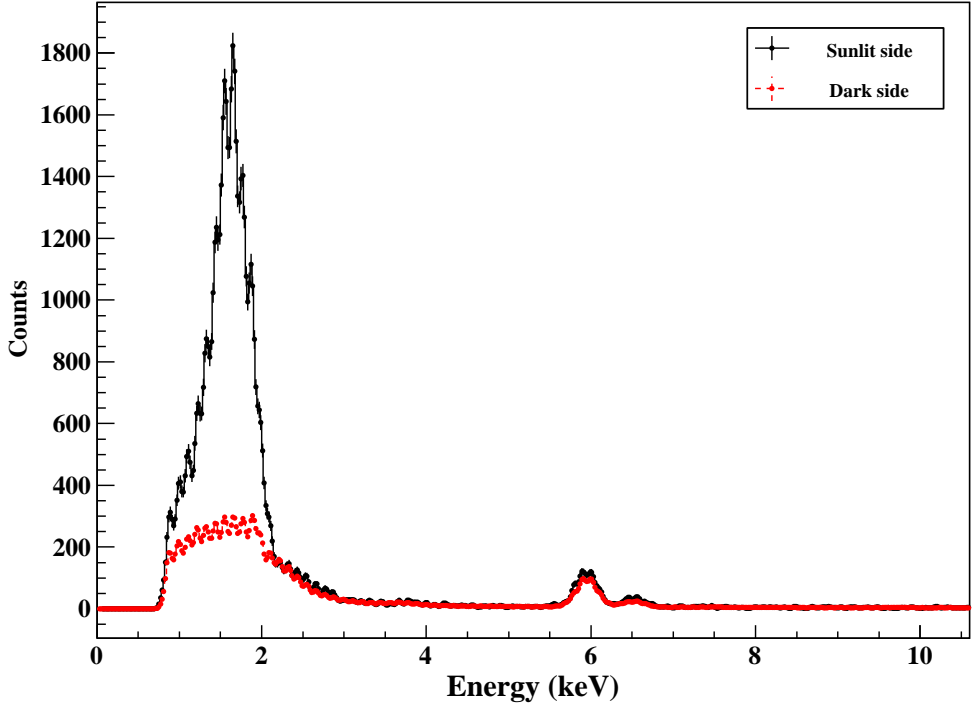}
   \caption{An example   spectra observed by CE2XRS in sunlit side (in black point) and  in the dark side with solar incident angle larger than $120^{\circ}$ (in red point)  when $T_{solar}=4 MK$.}
   \label{Fig_bg}
   \end{figure}

The X-rays detected by CE2XRS in orbit include not only the fluorescent X-rays from the  lunar surface, but also the  backgrounds which come from internal electronic noise of the instrument (\citealt{Peng+2009}),  external cosmic ray induced background and scattered solar X-rays (\citealt{Nittler+etal+2001}). \cite{Ban+etal+2014} made a comparison of the electronic noise background at different temperatures. It was found that  the electronic noise of the detector was fairly constant and change slightly with the detector's temperature. The overall shape of the cosmic ray induced background spectra also tends to be constant over the timescale of a few hours, but the magnitude varies (\citealt{Lim+Nittler+2009}; \citealt{Nittler+etal+2011}; \citealt{Narendranath+etal+2011}). In our analysis, the internal electric noise and cosmic ray induced background are  determined by  summing over all measurements in the same solar condition and in the dark side of the Moon. This method for background determination was widely adopted by many investigators (\citealt{Clark+Trombka+1997}; \citealt{Peng+2009}; \citealt{Nittler+etal+2011}; \citealt{Narendranath+etal+2011}; \citealt{Weider+etal+2012}; \citealt{Ban+etal+2014}). \cite{Ban+etal+2014} used the dark side with solar incident angle larger than $90^{\circ}$ as the background. To avoid illuminations of Sun light and zodiacal light into the detector directly near the terminator region, only the spectra with a solar X-ray incident angle greater than $120^{\circ}$ is added for background determination in our analysis. In our case, the background variation with incident angle larger than $120^{\circ}$ is more stable as a function of incident angle. And then, the background is normalized  according to the accumulation time and subtracted directely in the spectra acquired in the sunlit side of the Moon.  Figure~\ref{Fig_bg} shows an example  spectra observed by CE2XRS in sunlit side and in dark side with solar incident angle larger than $120^{\circ}$ when $T_{solar}=4 MK$. The scattered solar X-ray background in the sunlit side has been calculated theoretically by \cite{Clark+Trombka+1997}. \cite{Nittler+etal+2001} have used this method to predict scattered solar X-rays spectra at different solar plasma temperatures. The shape in low energy part of scattered solar X-ray spectra is approximately a Gaussian profile at low solar plasma temperature. Therefore we use a Gaussian function to model the spectra of scattered solar X-rays background. As shown in the following section, this function describes the low energy tail of the spectra well.

\subsection{Spectral Deconvolution and Elemental Abundance Mapping}
\begin{figure}
   \centering
  \includegraphics[width=14.5cm, angle=0]{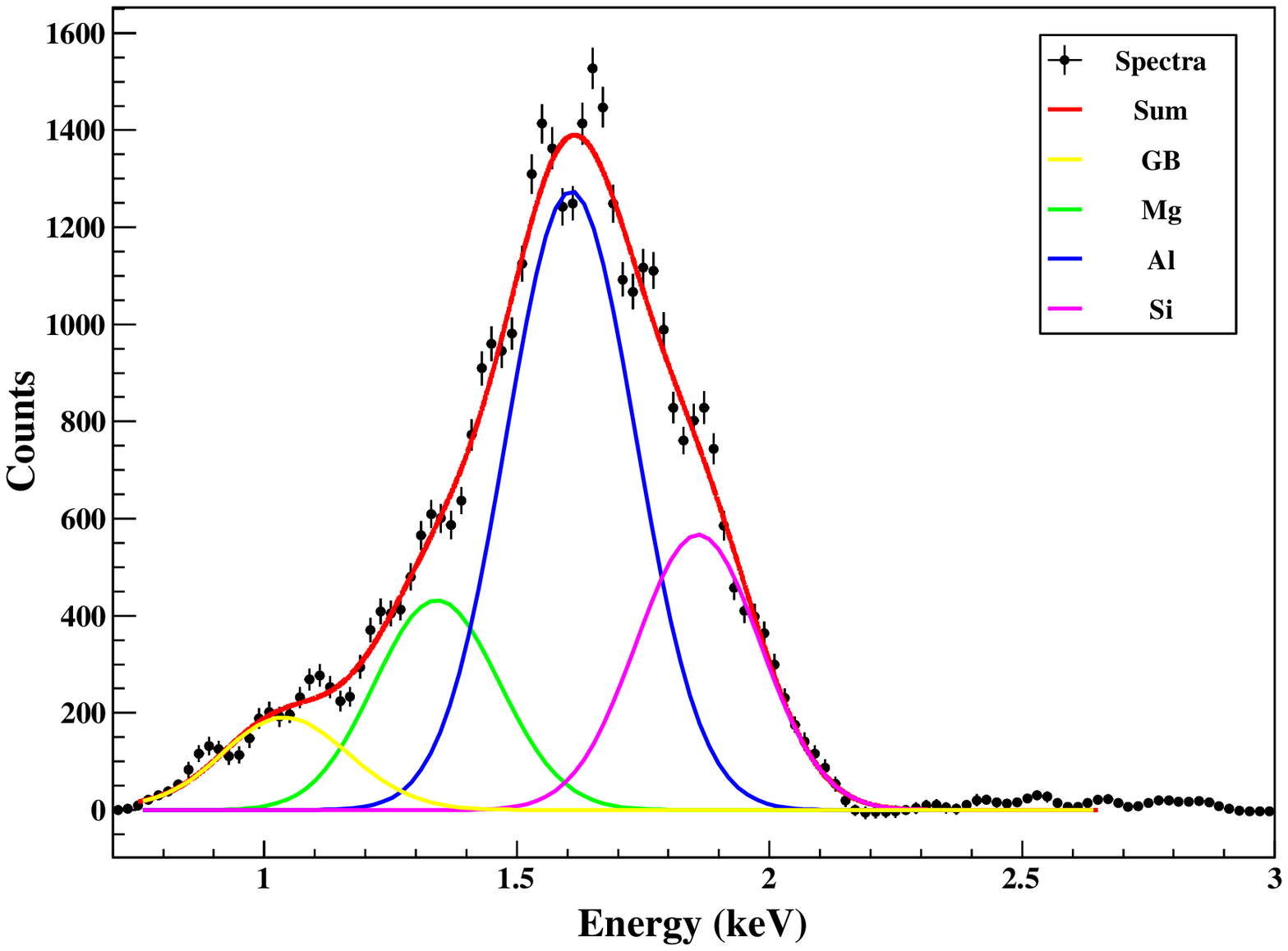}
   \caption{The spectra with electronic noise and cosmic induced background substracted are fitted by a sum of Gaussian functions for K$\alpha$ lines of Mg (green line), Al (blue line), and Si (pink line), and superimposed on a scatter solar X-ray background in yellow. The line in red is the best-fit result.}
   \label{Fig_fit}
   \end{figure}

  After the electronic noise and cosmic induced background subtracted, the residual spectra were deconvolved to obtain the flux of each major element, such as Mg, Al and Si. For most spectra, the  fitting function is  the sum of  three normalized Gaussian functions for Mg, Al, Si and a Gaussian function for scattered solar X-ray background (see Figure~\ref{Fig_fit}).    Each element intensity is determined by the best fit of the spectra with a minimum $\chi^2$ method that takes into account the error bar of each data point. The fitting procedure is coded and implemented in ROOT\footnote{The ROOT software can be freely downloaded from its official website https://root.cern.ch/drupal/} software.  One can also obtain the uncertainties of the X-ray flux of each element from the fitting results.  Based on the fitting results, we derived X-rays intensity ratio of each element relative to Si.  Si abundances only vary slightly on the lunar surface. In the analysis, ratioing to Si helps to remove matrix effects that come from compositional variations in the lunar regolith and from geometric corrections (\citealt{Clark+Trombka+1997}). Then, the elemental flux ratios are used to create global distribution maps on the lunar surface.

\section{Results And Discussions}
\label{sect:discussion}

\begin{figure}
   \centering
  \includegraphics[width=14.5cm, angle=0]{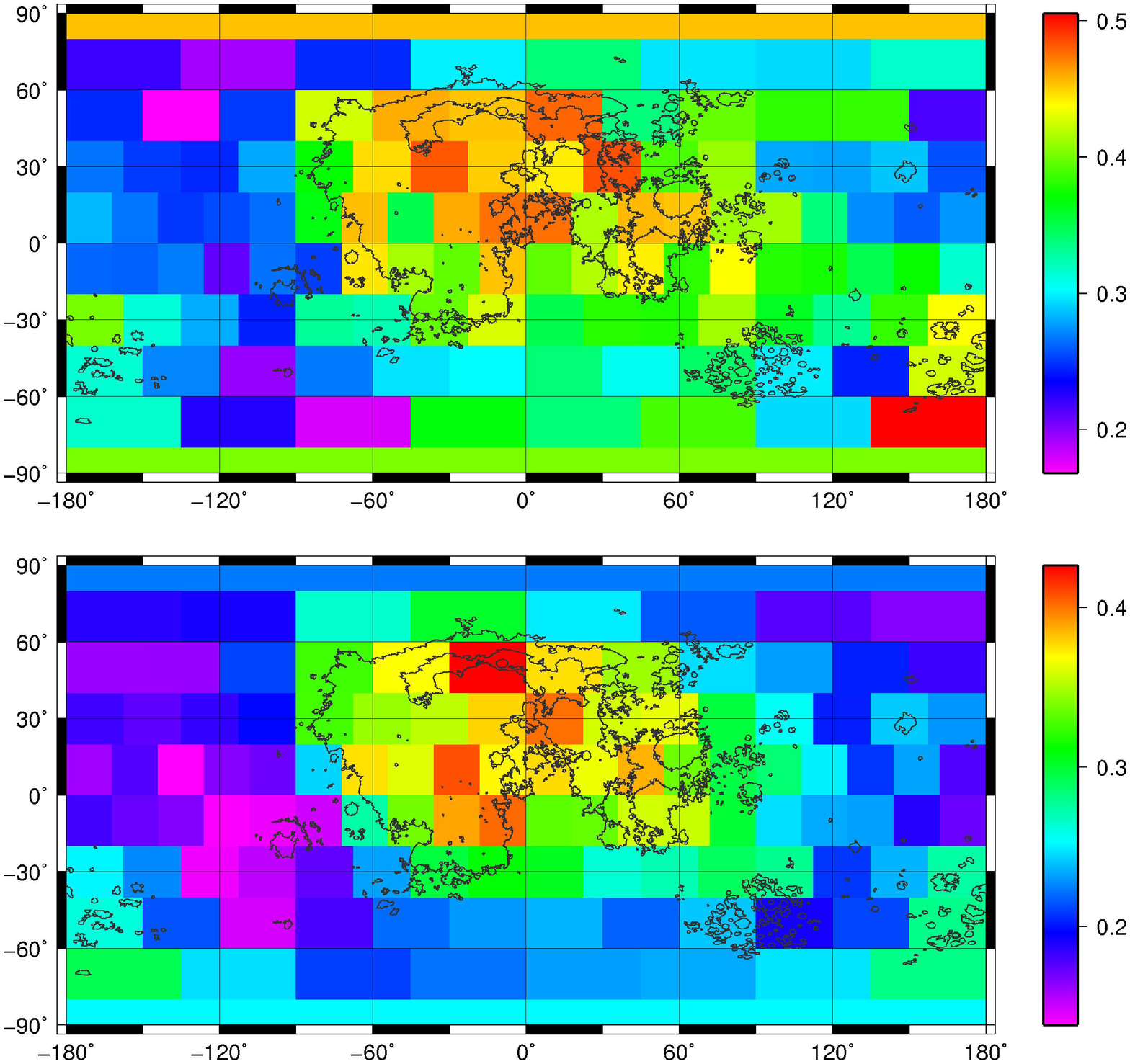}
   \caption{The equal-area ($20^\circ\times20^\circ$) global map of Mg/Si molar ratios distribution on lunar surface (Up: derived from CE2XRS; Down: derived from LPGRS).}
   \label{Fig_mg}
\end{figure}

\begin{figure}
   \centering
  \includegraphics[width=14.5cm, angle=0]{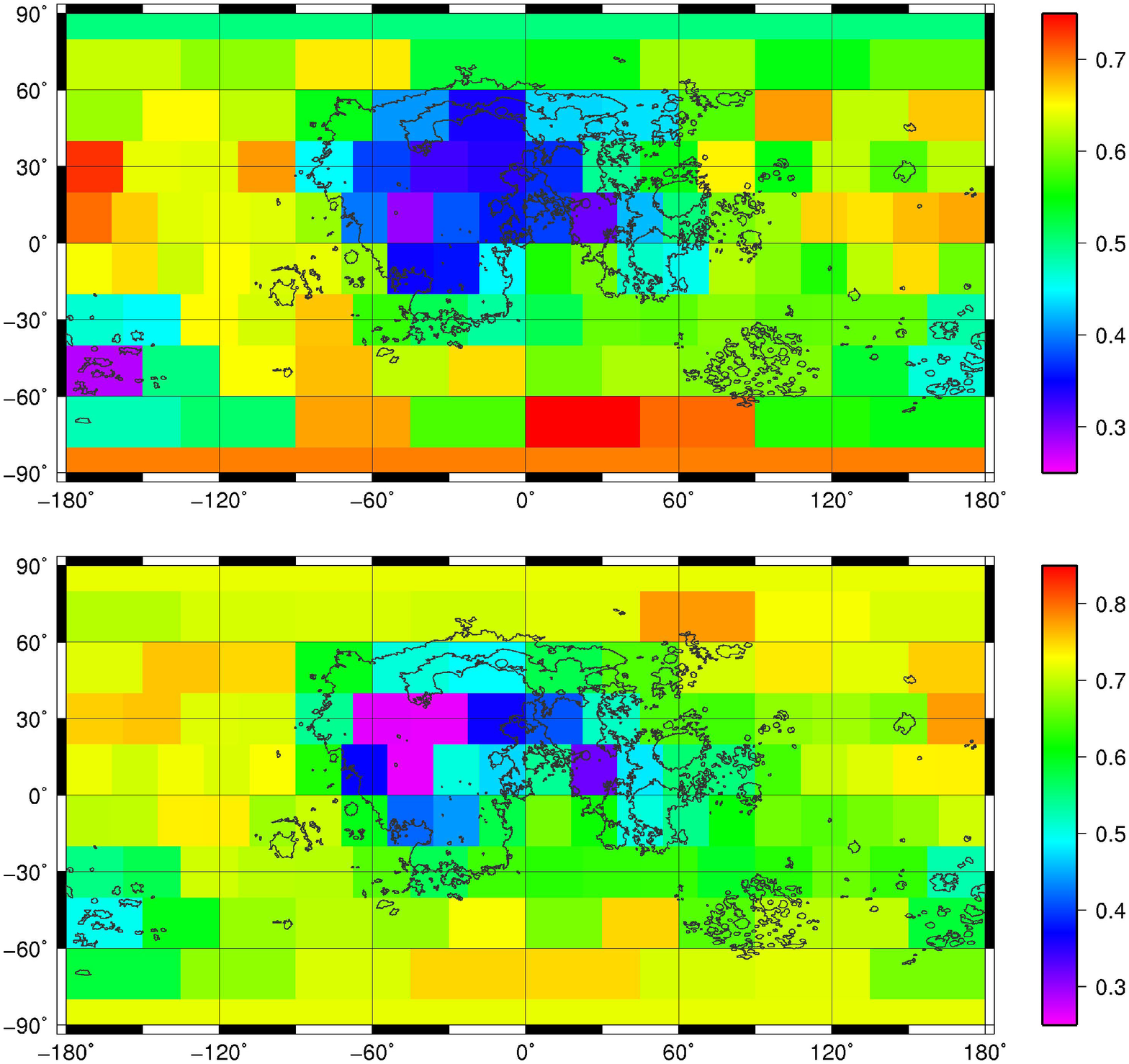}
   \caption{The equal-area ($20^\circ\times20^\circ$) global map of Al/Si molar ratios distribution on lunar surface (Up: derived from CE2XRS; Down: derived from LPGRS).}
   \label{Fig_al}
\end{figure}

After above data process steps are carried out, we  obtain the first global Mg/Si and Al/Si maps on the Moon derived from CE2XRS level-2C data on the condition that the solar temperature is 4 MK. The result is normalized by global Mg/Si and Al/Si data derived from Gamma Ray Spectrometer on Lunar Prospector (LPGRS) (\citealt{Prettyman+etal+2006}). Figure~\ref{Fig_mg} shows the global Mg/Si map with a resolution of $20^\circ\times20^\circ$ and the same resolution map from LPGRS. Figure~\ref{Fig_al} shows the global Al/Si map with a resolution of $20^\circ\times20^\circ$ and the same resolution map from LPGRS. From Figure~\ref{Fig_mg}, we can find that the high Mg/Si ratios are mainly concentrated at the PKT (Procellarum KREEP Terrane) and SPA (South Pole-Aitken) while FHT (Feldspathic Highlands Terrane) has a low value. However, the high Al/Si ratios are inversely concentrated at the FHT and with a low value at PKT and SPA, as shown in Figure~\ref{Fig_al}. All these maps are consistent well with the maps derived from LPGRS.

The possible reasons of the differences in certain regions of elemental abundance ratios derived between CE2XRS and LPGRS are presented as follows. The solar spectra used for CE2XRS derivation in initial analysis is from GOES data, which may result in uncertainty to some extent. The solar activity calibration model that use CE2XRS Solar X-ray Monitor data will be established in future work. On the other hand, the detection depths of the two techniques are different. X-ray fluorescence is exicited by incident solar X-rays at the upmost surface of the lunar regolith. The penetration depth of solar X-rays is $\sim$ 100 $\mu$m, while the penetration depth of gamma rays is on centimeter scale.

\section{CONCLUSIONS}
We have described data analysis procedures in detail for CE2XRS data and have presented the global distributions of Mg/Si and Al/Si on the lunar surface from CE2XRS, which are the first global maps derived from X-ray  spectroscopy. Three major geological units (lunar mare, highland and South Pole-Aitken basin) are identified. The results demonstrate that X-ray remote-sensing technique can also provide important compositional information about lunar surface.

\normalem
\begin{acknowledgements}
We thank Prof. Wang Huanyu, Dr. Peng Wenxi, Dr. Guo Dongya and Dr. Xiao Hong from the Institute of High Energy Physics, Chinese Academy of Sciences, for valuable discussions and help on data analysis. Special thanks go to the Ground Application System of Lunar Exploration, National Astronomical Observatories, Chinese Academy of Sciences for providing the CE2XRS data. This research is supported by the Science and Technology Development Fund of Macao (Grant Nos. 068/2011/A, 048/2012/A2, 039/2013/A2, 091/2013/A3 and 020/2014/A1) and by the Key Research Program of the Chinese Academy of Sciences (Grant No. KGZD-EW-603).
\end{acknowledgements}

\bibliographystyle{raa}
\bibliography{ms2096}

\end{document}